\documentclass[11pt,twoside]{article}
\usepackage{./asp2014}

\aspSuppressVolSlug
\resetcounters

\begin{document}

\title{The Star Formation Rate of Massive Dusty Galaxies at Early Cosmic Times}
\author{Zacharias E. Escalante,$^1$ Shardha Jogee,$^1$ and Sydney Sherman$^1$
\affil{$^1$Department of Astronomy, University of Texas at Austin, Austin, TX 78712-1205, USA}}

\paperauthor{Zacharias E. Escalante}{zescalante@utexas.edu}{ORCID_Or_Blank}{University of Texas at Austin}{Astronomy Department}{Austin}{TX}{78712-1205}{USA}
\paperauthor{Shardha Jogee}{sj@astro.as.utexas.edu}{ORCID_Or_Blank}{University of Texas at Austin}{Astronomy Department}{Austin}{TX}{78712-1205}{USA}
\paperauthor{Sydney Sherman}{ssherman@astro.as.utexas.edu}{ORCID_Or_Blank}{University of Texas at Austin}{Astronomy Department}{Austin}{TX}{78712-1205}{USA}

\begin{abstract}
We explore how the estimated star formation rate (SFR) of a sample of isolated, massive dusty star-forming galaxies at early cosmic epochs ($1.5 < z < 3.5$)  changes when their ultraviolet (UV) to near-infrared (NIR) spectral energy distribution is extended to longer wavelengths by adding far-infrared/sub-millimeter data to trace the reprocessed radiation from dust heated by young massive stars. We use large-area surveys with multi-wavelength datasets that include DECam UV-to-optical, 
VICS82 NIR, \textit{Spitzer}-IRAC NIR, and \textit{Herschel}-SPIRE far-infrared/sub-millimeter data.   We find that the inclusion of far-infrared/sub-millimeter data leads to  SFRs that span  $\sim$100-3500 $M_{\odot} yr^{-1}$  and are higher than the extinction-corrected UV-based SFR by an average factor of $\sim$3.5, and by a factor of over 10  in many individual galaxies. Our  study demonstrates the importance of far-IR/sub-millimeter data for deriving accurate SFRs in  massive dusty galaxies at early epochs, and underscores the need for next-generation far-IR/sub-millimeter facilities with high sensitivity, field of view, and angular resolution.
\end{abstract}

\section{Introduction}

In the early Universe, many young galaxies possess a rich supply of gas and dust, and are actively forming stars.  Some of the most massive dusty star-forming galaxies (DSFGs) have been found to show extremely high star formation rates (SFRs) above 1000 $M_{\odot} yr^{-1}$ (e.g., \cite{Chapman2005}, \cite{Ivison2013}, \cite{Casey2014} and references therein).  In the star-forming regions of a galaxy, young hot massive stars emit most of their energy at rest-frame ultraviolet (UV) and optical wavelengths. In a DSFG, a small portion of this UV and optical light reaches the observer, while a dominant fraction is absorbed by dust and subsequently re-emitted by the heated dust at longer far-infrared/sub-millimeter wavelengths. In order to accurately estimate the SFRs of  DSFGs, we must map their spectral energy distribution (SED), not only at short wavelengths, but also at the longer far-infrared/sub-millimeter wavelengths that trace the emission from dust heated by hot massive stars. 

Unfortunately, many large-area surveys of galaxies only have high angular resolution data at shorter (UV to NIR) wavelengths, and studies based on these surveys estimate the SFR from this limited dataset. In order to at least partially correct for the fact that some of the UV light is absorbed by dust, such studies 
apply an extinction correction based on the UV-to-NIR SED and derive an extinction-corrected UV-based SFR (SFR$_{UV-Corrected}$).  Despite this extinction correction, this measure of SFR may still  underestimate the true SFR in massive DSFGs.  In this work, we explore how the extinction-corrected UV-based SFR (SFR$_{UV-Corrected}$) estimated from the UV-to-NIR SED compares to the SFR (SFR$_{UV-FIR}$) estimated after the SED is extended to longer wavelengths through the addition of far-infrared/sub-millimeter data.  We focus on  massive DSFGs at redshifts of  $1.5 < z < 3.5$,  corresponding to  the important epoch  when the Universe was only $\sim$30\% to 15\% of its present age and the cosmic SFR density peaked. Throughout this work we adopt a flat $\Lambda$CDM cosmology with $h = 0.7$, $\Omega_m = 0.3$, and $\Omega_{\Lambda} = 0.7$.

\section{Data and Sample}

Our study focuses on a sample of galaxies drawn from a  large-area (17.2 $deg^2$ in SDSS Stripe 82) multi-wavelength survey that was conducted to study massive galaxies at $1.5 < z < 3.5$.  The survey photometric datasets include 
UV-to-optical  \textit{u,g,r,i,z}  data from the Dark Energy Camera (DECam)  \citep{Wold2019}, 
NIR  $J$ and $K_s$ data from the VICS82 Survey \citep{Geach2017}, 
NIR 3.6 and 4.5 $\mu$m data from \textit{Spitzer}-IRAC \citep{Papovich2016},
far-IR/sub-millimeter data at 250, 350, and $500 \  \micron$ from  \textit{Herschel}-SPIRE (HerS, \citealt{Viero2014}), and 
XMM-Newton and Chandra X-ray Observatory X-ray data from the Stripe 82X survey (\citealt{LaMassa2013a}, \citealt{LaMassa2013b}, \citealt{Ananna2017}).

Our study aims to target massive DSFGs at $1.5 < z < 3.5$. To achieve a clean sample of these objects, we applied the following selection criteria to a set of massive star-forming galaxies selected by \cite{Sherman2019}. First, a group of 5,352 galaxies were filtered from the complete set to have photometric redshifts in the range $1.5 < z < 3.5$, stellar masses $M_{\star}/M_{\odot} \geq 10^{11}$, and specific star formation rate (sSFR)  $> 10^{-11}$ $yr^{-1}$. These galaxies were required to contain signal-to-noise S/N $\geq$ 5 in the IRAC 3.6 and 4.5 $\micron$ filters and the DECam r-band filter, to ensure a sample of galaxies that are both massive and star-forming (see \cite{Sherman2019} for a detailed discussion of these criteria).

Next, the set of 5,352 galaxies were position matched to the \textit{Herschel}-SPIRE  (HerS) data. We note here that  the 
HerS 250, 350, and $500 \  \micron$ data have very poor angular resolution (full width half max values of $\sim$18, 24, and 35 arcseconds, respectively), compared to the angular resolution (1 to 2 arcseconds)
of the UV-to-NIR data. At  $1.5 < z < 3.5$, in high-density environments where galaxies are crowded together and separated by only a few arcseconds, the low resolution \textit{Herschel}-SPIRE images can blend several galaxies into one single far-IR/sub-millimeter source. To avoid the issue of blended sources in the HerS data, our study is compelled to focus only on isolated galaxies.

The isolated galaxies were identified through the following steps.  Each of the 5,352 massive star-forming galaxies at $1.5 < z < 3.5$ was assumed to have a potential HerS match if its r-band position was within 4 arcseconds of the HerS  250 $\micron$ source position. 186 galaxies had a position match and were found to have non-zero flux in all three HerS bands. These 186 galaxies were then visually inspected to identify isolated galaxies, which were defined to be systems  whose DECam r-band position did not have another object within 9 arcseconds.  This process yielded a sample of 38 isolated, massive star-forming galaxies with \textit{Herschel}-SPIRE  250, 350, and $500 \  \micron$ data.

Finally, the positions of these 38 isolated, massive star-forming galaxies were cross-matched with positions of objects in the Stripe 82X X-ray catalog in order to determine whether any of these galaxies host  X-ray luminous active galactic nuclei (AGN). One galaxy was found to contain such an AGN, and was thus excluded from further analysis. The final sample of 37 isolated  galaxies with \textit{Herschel}-SPIRE far-IR/sub-millimeter data constitutes only  $0.69\%$ of the massive star-forming galaxy sample (5,352 galaxies) in this region.

\section{Methodology}

We measure the SFR and stellar mass of our sample of 37 isolated, massive DSFGs by first constructing two empirical SEDs for each galaxy. The first empirical SED covers UV-to-NIR wavelengths and uses nine photometric points from the DECam \emph{u,g,r,i,z,} VICS82 $J$ and $K_s$, and \textit{Spitzer}-IRAC (3.6 and 4.5 $\micron$) data.  The second empirical SED includes the latter nine data points and an additional three \textit{Herschel}-SPIRE  250, 350, and $500 \  \micron$ data points, thus covering UV to far-infrared/sub-millimeter wavelengths.

The first empirical UV-to-NIR SED is fitted with the  EAZY-py SED-fitting code to derive the SFR and stellar mass of each galaxy. EAZY-py is a Python-based version of EAZY (\cite{Brammer2008}), optimized for fitting SED data from UV to NIR wavelengths. This code makes use of a non-negative linear combination of twelve Flexible Stellar Population Synthesis (FSPS) templates (\cite{Conroy2009}, \cite{Conroy2010}) to create a best fit model SED for an object's flux values using a $\chi^2$ minimization routine. Through this method of linearly-combined template fitting, EAZY-py fits the UV-to-NIR empirical SED with a model SED. Although the model SED from EAZY-py extends from UV out to far-IR/sub-millimeter wavelengths, the model is primarily constrained by data at wavelengths less than  5 $\micron$ (Figure \ref{EAZY-py}).

\articlefigure{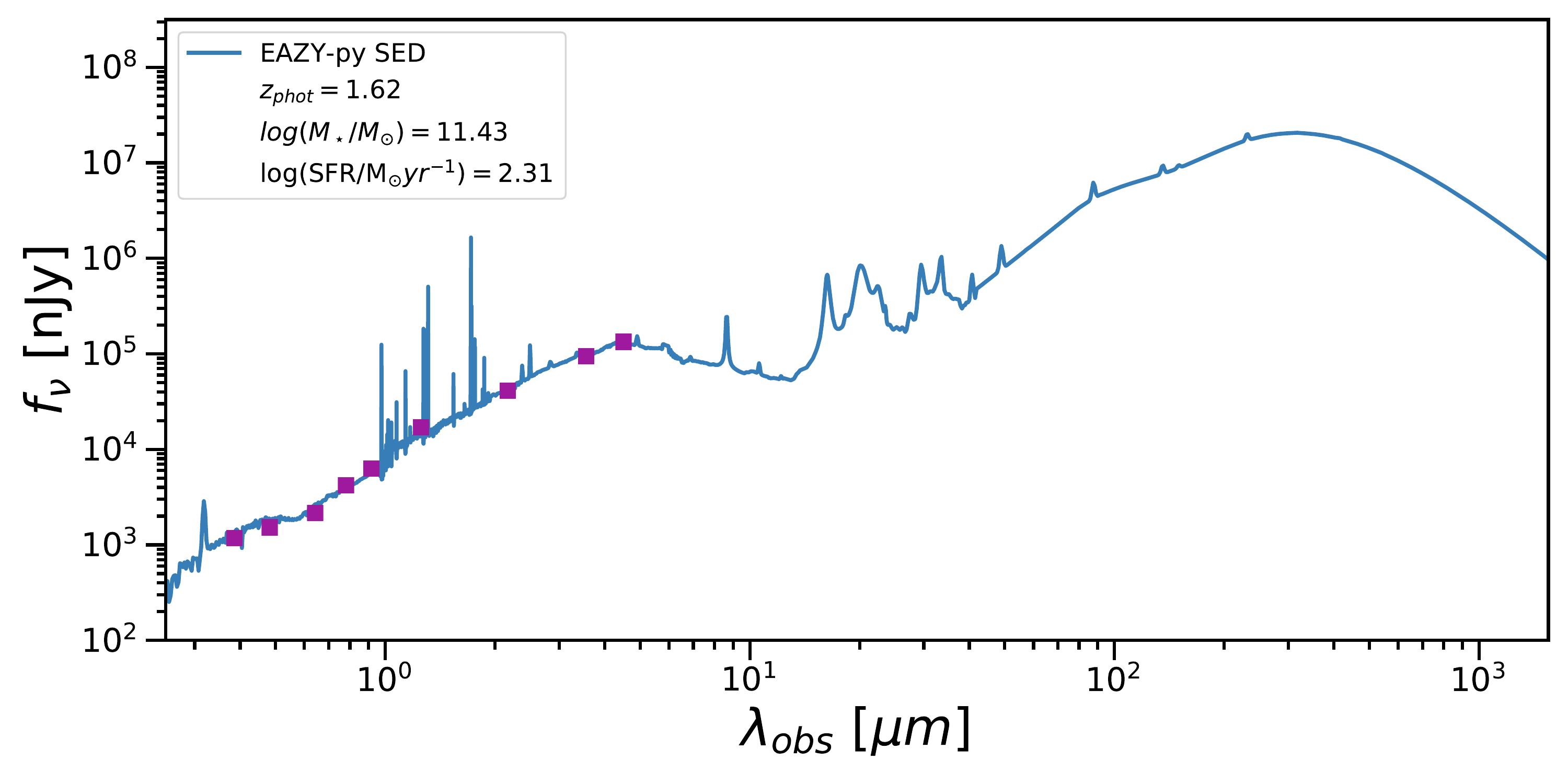}{EAZY-py}{Best fit model SED from EAZY-py for a galaxy at $z_{phot} = 1.62$.
EAZY-py produces a best fit model SED (blue curve) to the empirical SED, which covers UV to NIR wavelengths from nine photometric points (purple squares) based on the DECam \emph{u,g,r,i,z,} VICS82 $J$ and $K_s$, and \textit{Spitzer}-IRAC 3.6 $\micron$ and 4.5 $\micron$ data. Error bars are plotted for all photometric points and may be smaller than the symbol.}

\articlefigure{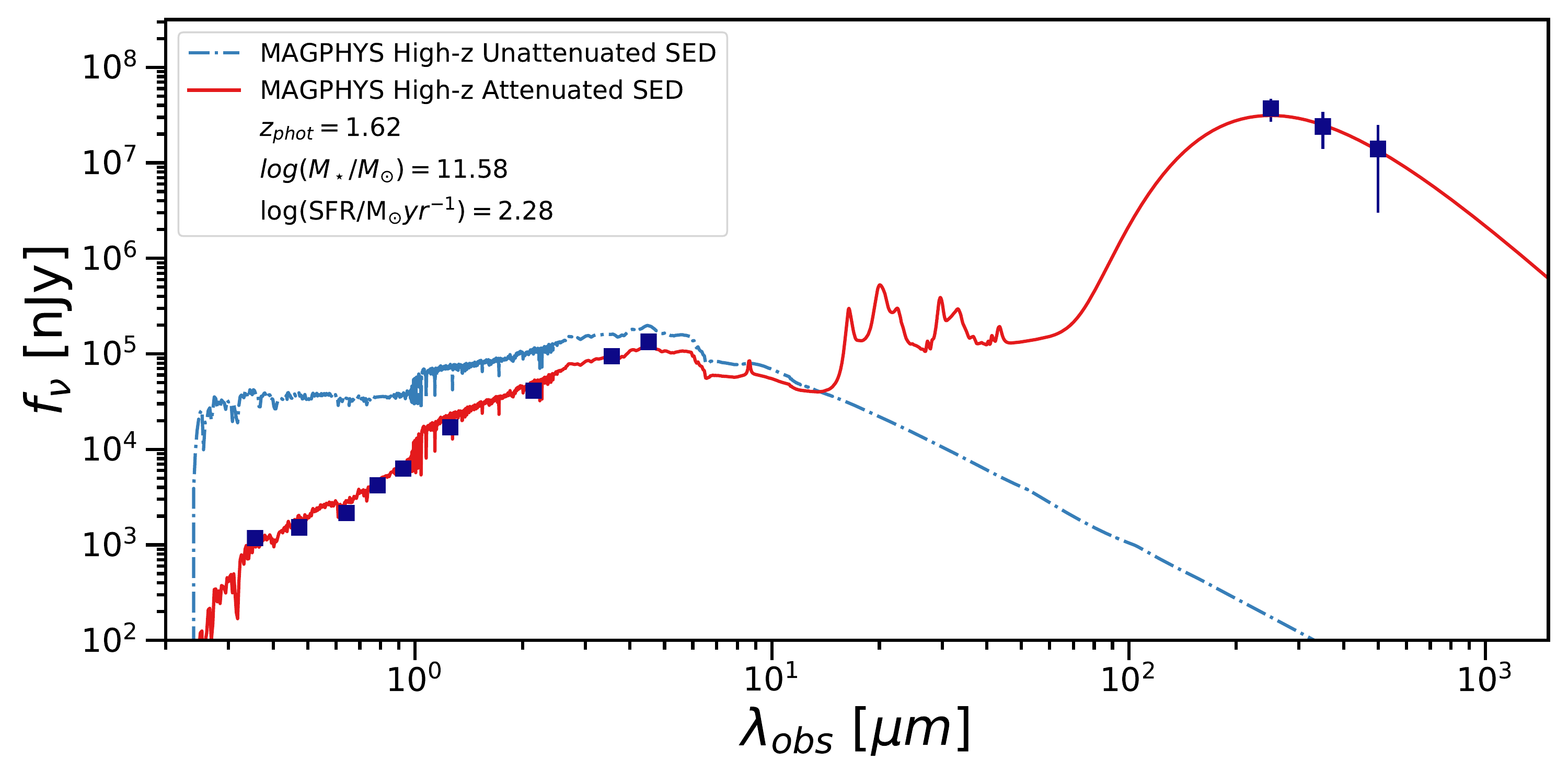}{MAGPHYS}{Best fit model SED from MAGPHYS High-z for the same galaxy as in Figure \ref{EAZY-py}. The empirical SED shown here is built by taking the empirical UV-to-NIR SED in Figure \ref{EAZY-py} and adding the three \textit{Herschel}-SPIRE  250, 350, and $500 \  \micron$ data points, thereby extending the coverage to far-IR/sub-millimeter wavelengths where emission from dust heated by massive stars is detected. In the best fit model SED, the blue dashed curve represents the inferred unattenuated flux from stellar emission, while the solid red curve represents the attenuated flux from stellar and dust emissions.}

The second empirical SED, with twelve data points covering UV to far-IR/sub-millimeter wavelengths, is fitted with a different SED-fitting code known as MAGPHYS High-z (Multi-wavelength Analysis of Galaxy Physical Properties; \cite{daCunha2008}, \cite{daCunha2015}). We do not use EAZY-Py to fit the second empirical SED  because it assigns very low weights to the \textit{Herschel} far-IR/sub-millimeter data due to their signal-to-noise being much lower than that of the UV-to-NIR data. The MAGPHYS High-z code uses theoretical SED templates to determine the best fitting SED for each object. The model libraries in the High-z code -- compared to the original MAGPHYS code (\cite{daCunha2008}) -- are created with highly redshifted ($z \textgreater 1$) galaxy properties in mind. In particular, models with higher dust optical depths and star formation rates, as well as younger ages are included. 

With a starting library of over 50,000 stellar population and dust emission spectra of mock galaxies, MAGPHYS High-z redshifts each model spectra to match the given value of redshift. Since MAGPHYS High-z requires a prior redshift for each galaxy to begin the fitting procedure, the photometric redshifts calculated from EAZY-py were used (our sample of 37 selection galaxies have redshifts spanning $1.5<z_{phot}<3.27$). Each model is then compared to the observed object fluxes using a Bayesian analysis method, calculating the $\chi^2$ value and creating a likelihood distribution of each galaxy parameter. The 50th percentile of the distribution is then chosen as the best fit value of the parameter, with the 16th and 84th percentile values taken as the lower and upper error estimates, respectively. It is important to note that MAGPHYS High-z fits the rest-frame NIR to far-IR part of the spectrum separately from the rest of the data. 
Thus, MAGPHYS High-z provides a means of properly utilizing both the UV-to-NIR and the \textit{Herschel} far-IR/sub-millimeter data to infer the SFR (SFR$_{UV-FIR}$) (Figure \ref{MAGPHYS}).

\section{Results and Conclusion}

\articlefigure{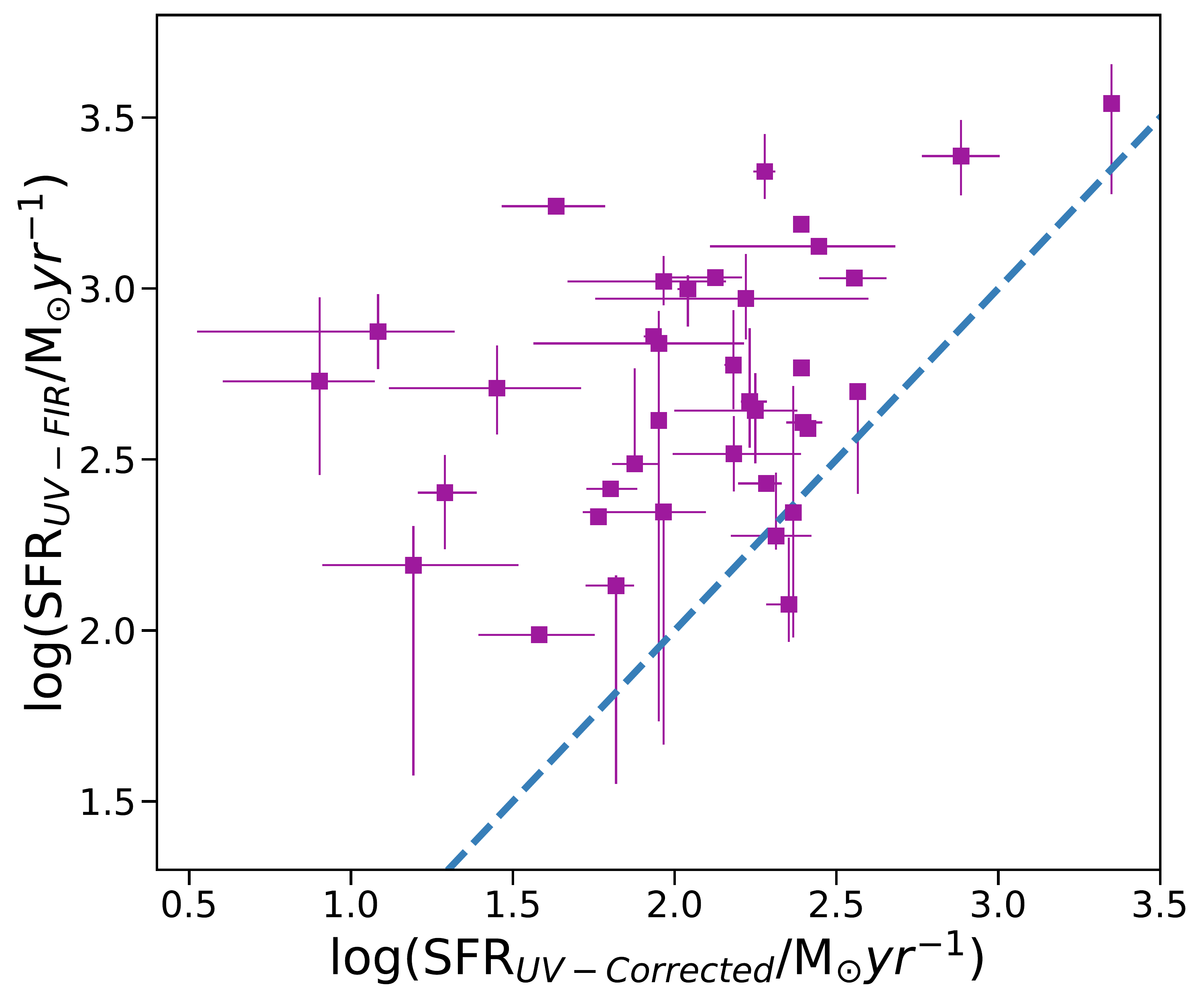}{SFR}{Comparison between the extinction-corrected UV-based SFR (SFR$_{UV-Corrected}$) estimated from the empirical UV-to-NIR SED and the SFR (SFR$_{UV-FIR}$) estimated from the empirical SED that extends from UV to far-infrared/sub-millimeter wavelengths. We find that SFR$_{UV-FIR}$ is higher than SFR$_{UV-Corrected}$ by an average factor of $\sim$3.5  and by a factor of over 10 in seven galaxies ($\sim$$19\%$ of our sample).  
}

\articlefigure{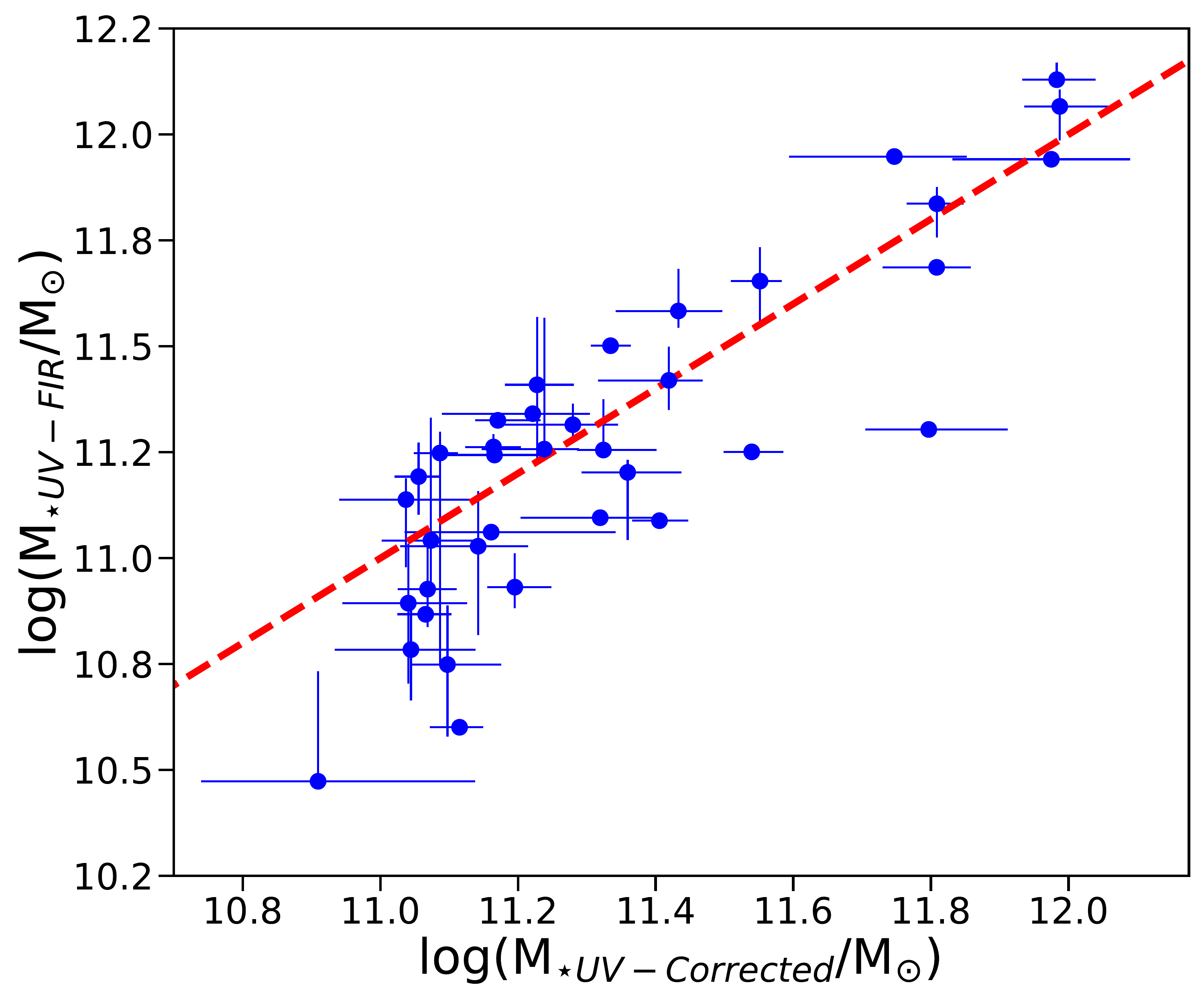}{Mass}{Same as in  Figure \ref{SFR}, but for galaxy stellar masses. Estimates of galaxy stellar masses at $1.5 < z < 3.5$ are not significantly impacted by the inclusion of  far-infrared/sub-millimeter data.}

Figure \ref{SFR} compares the extinction-corrected UV-based SFR (SFR$_{UV-Corrected}$) derived from the empirical SED at short (UV-to-NIR) wavelengths to the SFR (SFR$_{UV-FIR}$) derived from the empirical SED that extends from the UV to far-infrared/sub-millimeter wavelengths. Figure \ref{Mass} shows the equivalent comparison for the galaxy stellar mass derived from the
two SEDs.

It is clear from Figure \ref{Mass}  that estimates of galaxy stellar masses at $1.5 < z < 3.5$ are not significantly impacted by the inclusion of  far-infrared/sub-millimeter data.   This result is expected, given that stellar masses are primarily traced by rest-frame NIR light or rest-frame red optical light at these epochs.

Conversely, Figure \ref{SFR} shows that the  inclusion of the far-infrared/sub-millimeter data leads to SFRs (SFR$_{UV-FIR}$) that are higher than the extinction-corrected UV-based SFR (SFR$_{UV-Corrected}$) by an average factor of $\sim$3.5 and by a factor of over 10 in seven galaxies ($\sim$$19\%$ of our sample). This result shows that the inclusion of far-infrared/sub-millimeter data is extremely important for deriving accurate SFRs in massive dusty galaxies at early epochs. The UV-to-FIR based SFRs (SFR$_{UV-FIR}$)  are quite high, ranging from $\sim$100 to 3500  $M_{\odot} yr^{-1}$. How does our sample of massive, isolated DSFGs at $1.5 < z < 3.5$ compare to the general massive galaxy population at these epochs?
We note from Section 2 that our sample represents only $0.69\%$ of the general starting population of 5,352 massive star-forming galaxies at $1.5 < z < 3.5$.  Furthermore, the UV-to-FIR based SFRs (SFR$_{UV-FIR}$)  of our sample of galaxies lie well above the main sequence at $1.5<z<3.5$, generally by over 1 dex. These considerations suggest that these DSFGs represent extreme galaxies in terms of star-formation activity at these epochs. 

At the present time, the low sensitivity and angular resolution of far-infrared/sub-millimeter data over large survey areas 
allow us to include such data in the analysis of only the brightest and isolated DSFGs. In order to accurately estimate the obscured SFR for a large number of representative galaxies, we require next-generation far-infrared/sub-millimeter facilities with high sensitivity, angular resolution, and  field of view.

\acknowledgements The authors gratefully acknowledge support from the University of Texas at Austin, NSF grants AST 1614798 and 1413652, as well as NSF/DoD REU grant AST 1757983. 


\bibliography{citations}

\end{document}